\documentclass[dvips]{article}

\usepackage{icrc2011}

\title{Detailed Multifrequency Study of a Rapid VHE Flare of Mrk501 in May 2009}
\newcommand{\etal}{\MakeLowercase{\textit{et al. }}} % "et al."
\shorttitle{A. Pichel \etal Mrk 501 VHE flare}

\authors{A. Pichel$^{1}$ for the VERITAS Collaboration$^{2}$, D. Paneque$^{3}$ for the {\it Fermi}-LAT Collaboration and the others groups/Instruments that participated in the Mrk 501 campaign on 2009}
\afiliations{$^1$Instituto de Astronom\'ia y F\'isica del Espacio, Argentina\\ 
$^2$see J. Holder et al. (these proceedings) or http://veritas.sao.arizona.edu/conferences/authors?icrc2011 \\
$^3$Max Planck Institute for Physics, Germany}
\email{anapichel@iafe.uba.ar}

\abstract{We present observations of the gamma-ray blazar Markarian 501 between April 17 and May 5, 2009, with the Whipple 10-m telescope, VERITAS, the {\it Fermi} Large Area Telescope (LAT), {\it Swift} and {\it RXTE} as part of a 4.5-month multi-wavelength campaign. The presentation will focus on the strong very high energy (VHE) gamma-ray activity detected on May 1st with Whipple and VERITAS, when the measured flux ({\it E} $>$ 400 GeV) reached five times the flux of the Crab Nebula, coincident with an increase in the optical polarization by a factor of 5, and a rotation of the polarization angle by 15 degrees. We also show that, during this 3-week period, the largest flux and spectral variability is seen at the highest energies of the broadband spectral energy distribution.} 


\begin{document}
\maketitle

%Begin the section.

\vspace{-0.5cm}
\section{Introduction}
\vspace{-0.3cm}

Blazars are a subclass of active galactic nuclei (AGN) with their relativistic jet pointing close to the line of sight. These objects are among the most powerful astrophysical sources known. They exhibit rapid and irregular variability, with extreme outbursts from minutes to days in all energy bands. The fast events at high energy are thought to be produced in a compact internal zone close to the system (e.g. \cite{lab21}). 

For most blazars, the spectral energy distribution (SED) in a $\nu F_{\nu}$ representation shows a double-peaked structure, where the first peak, from radio to X-ray, is attributed to synchrotron radiation and the second peak, in the GeV to TeV, to inverse-Compton or hadronic processes.  

Mrk 501 is a nearby  $\gamma$-ray blazar (z = 0.034) first detected at TeV energies with the Whipple Telescope in 1996 \cite{lab1}. Since then, several multiwavelength campaigns were undertaken, mostly during episodes of high activity at long timescales or when flares on short timescales show up in the X-ray or $\gamma$-ray bands (e.g. \cite{lab2}, \cite{lab3}). In 1997, the source presented the most intense episode of high activity detected at very high energies by the Whipple telescope reaching a level of 10 times the Crab Nebula flux, and it was observed in X-rays by $BeppoSAX$ (\cite{lab4}, \cite{lab5}). 

\vspace{-0.5cm}
\section{Observations}
\vspace{-0.3cm}

We present here observations of Mrk 501 from April 17 to May 5, 2009, taken as part of a large-scale multi-wavelength campaign \cite{lab6}, which included a number of ground and space-based experiments covering the spectrum from radio to very high energy (VHE; {\it E} $>$ 100 GeV) $\gamma$-rays.  

In this 3-week time interval, Mrk 501 was observed at the VHE with the Whipple Telescope \cite{lab7} every night for a total of 20 hours and with VERITAS \cite{lab8,lab9} for a total of 4 hours \cite{lab10}. The VERITAS observations during the nights of April 30 and May 1 were made with two telescopes, those on the remaining nights with three telescopes, because the other units of the full four-telescope array were affected by hardware issues. At the high energy (HE; E $>$ 0.1 GeV) band, {\it Fermi}-LAT \cite{lab11}  monitored the source constantly, while the X-ray band was covered by {\it Swift}-XRT and RXTE-PCA. In optical and radio frequencies the source was covered by various instruments. In these proceedings we only show data from GASP, Mitsume and {\it Swift}-UVOT at optical frequencies, and Mets\"ahovi and OVRO at radio frequencies. More details of the instruments involved in the campaign and of the data analysis will be presented elsewhere.

\begin{figure*}[!ht]
   \centerline{
             \includegraphics[width=3.5in]{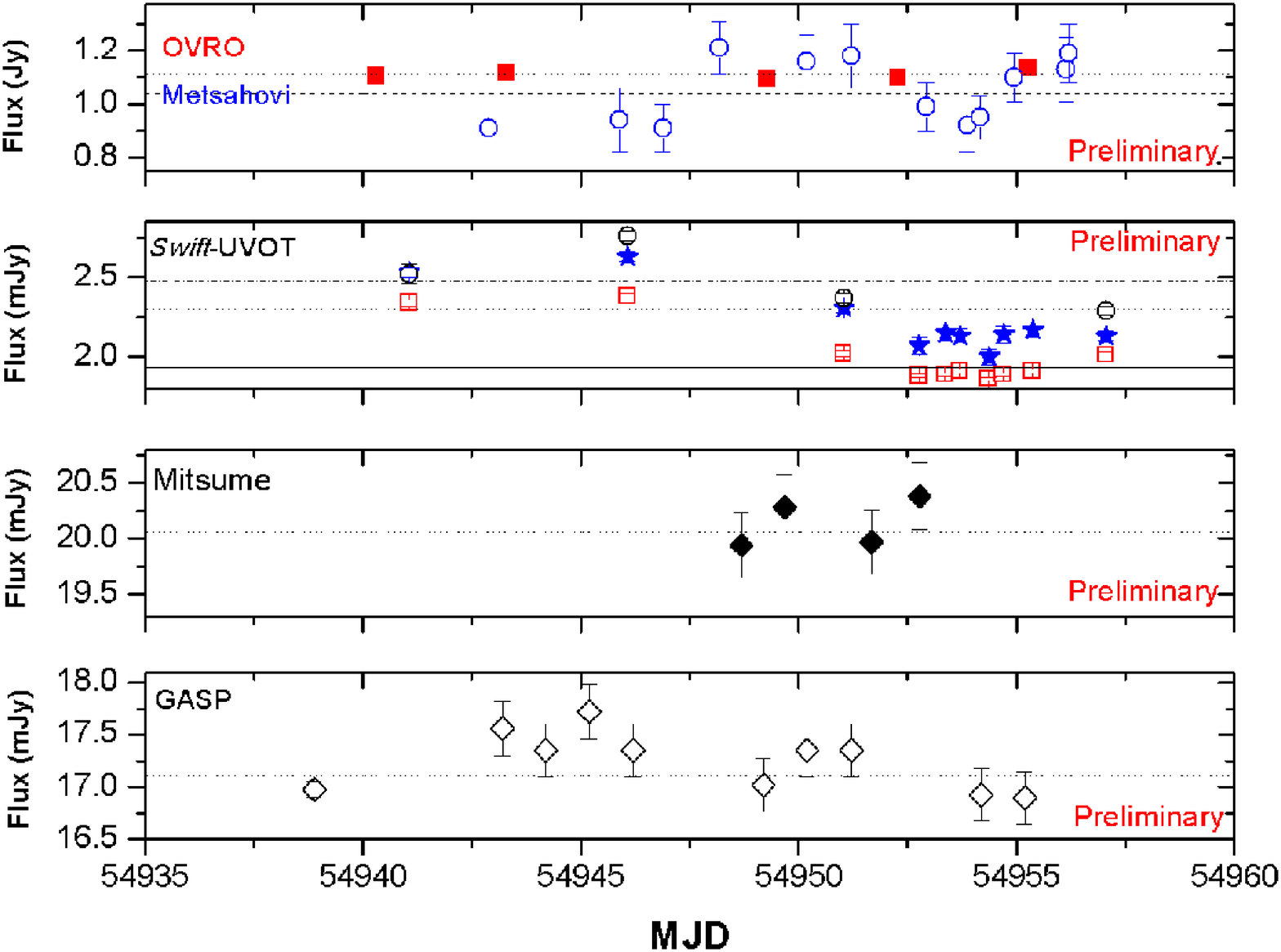}\label{fig1a}
              \hfil
              \includegraphics[width=3.5in]{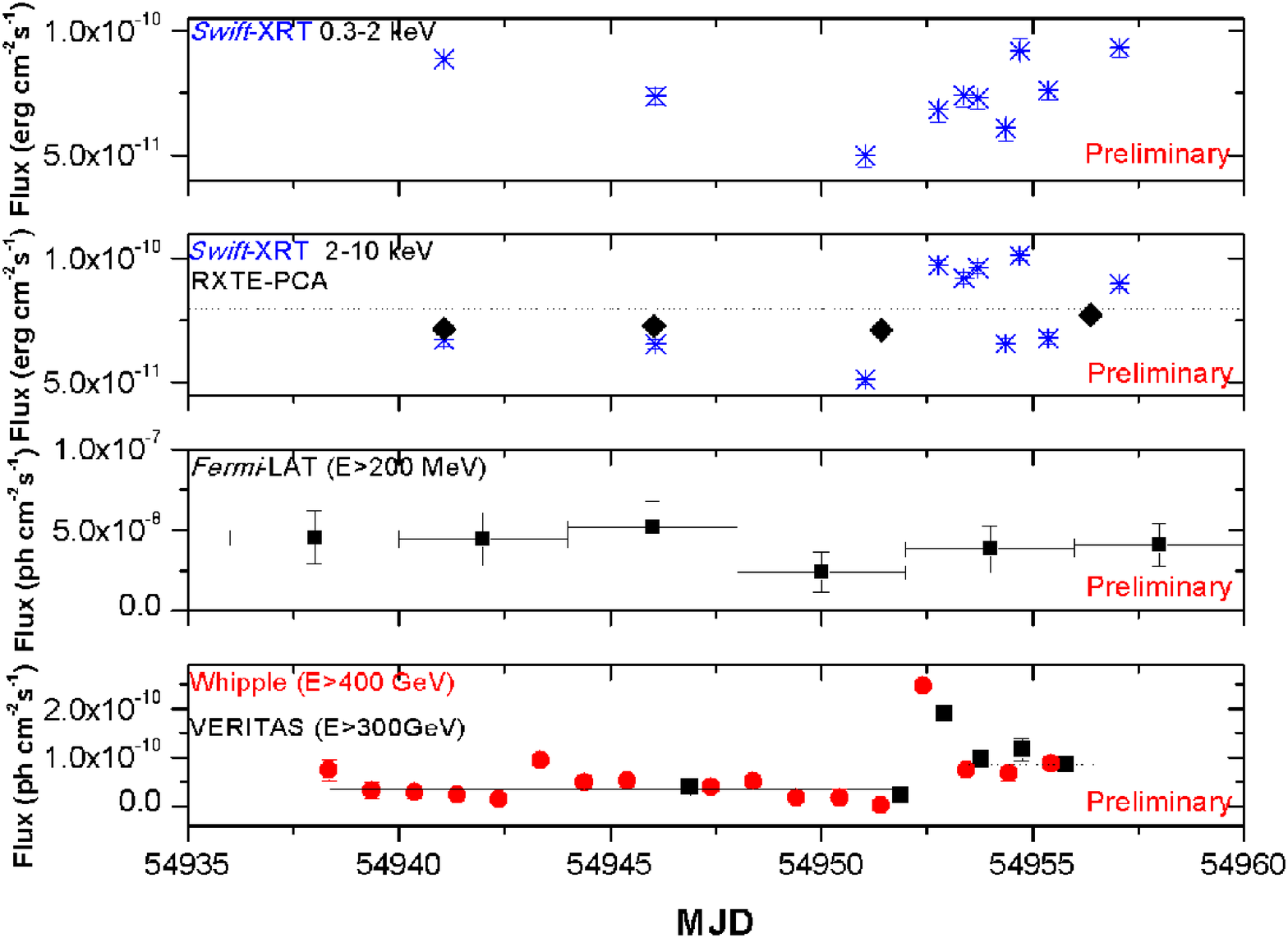} \label{fig1b}
             }
   \caption{Light curves for Mrk 501 from April 17 to May 5, 2009. Each dotted horizontal line represents a constant line fit for each instrument involved. {\sl \bfseries Left}: {\sl Top}: OVRO at 15 GHz (filled squares) and Mets\"ahovi at 37 GHz (open circles); {\sl Second}: {\it Swift}-UVOT in the Ultraviolet, with three different bands, UVW1 (260nm, open circles), UVM2 (220nm, filled stars) and UVW2 (193nm, open squares); {\sl Third}: Mitsume in g band (nightly average);  {\sl Bottom}: GASP in the R band (nightly average). {\sl \bfseries Right}: {\sl Top}: X-ray: {\it Swift}-XRT 0.3-2 keV (nightly average); {\sl Second}: X-ray: RXTE-PCA (filled diamonds) and {\it Swift}-XRT (crosses) 2-10 keV (nightly average); {\sl Third}: HE gamma-ray: {\it Fermi}-LAT (E $>$ 300 MeV; 5-day average); {\sl Bottom}: VHE $\gamma$-rays: Whipple ({\it E} $>$ 400 GeV,  normalised to {\it E} $>$300 GeV according to a power law with photon index -2.5; filled circles) and  VERITAS ({\it E} $>$ 300 GeV; filled squares; nightly average).}
   \label{double_fig}
 \end{figure*}

\vspace{-0.5cm}
\section{Results and discussion}
\vspace{-0.4cm}
\subsection{Light curves}
\vspace{-0.2cm}

Figure 1 shows the light curves for some of the instruments involved in the campaign. In radio and optical bands, the measured fluxes are constant (within statistical errors). In the X-ray bands one can see statistically significant variability (due to the high sensitivity of {\it Swift}-XRT and RXTE-PCA), but the flux variations are relatively small (well below 50 $\%$). On the other hand, in the VHE domain, VERITAS and especially Whipple measured (statistically significant) flux variations of a factor of a few during this time period and in particular up to a factor of 10 during MJD 54952. Therefore, during the above mentioned 3-week time interval, we find the highest variability at the highest energies.

At VHE, the light curve is consistent with the constant emission of the source ($3.9{\times 10^{-11}}~{\rm ph~cm^{-2}~s^{-1}}$) until the night of May 1 (MJD 54952), when a high-emission state was detected with Whipple and VERITAS, reaching a maximum $\gamma$-ray flux 10 times the average baseline flux, which is approximately 5 times the Crab Nebula flux. 

Figure 2 shows the Whipple 10-m and VERITAS light curves for May 1, 2009 where each point corresponds to a 4-minute bin. The flux increased by a factor of $\sim$5 in the first 30 minutes. In the days after the flare (MJD 54953-55), the source continued in a high state, and the flux each night was about twice the baseline flux.

The {\it Swift}-XRT observations are not strictly simultaneous to the VHE observations performed by Whipple and VERITAS. The time difference between {\it Swift} observations and Whipple and VERITAS observations is 7 hours.
Given that this source is known to have very fast variability (e.g., see \cite{lab2}), not having strictly simultaneous observations is a caveat that cannot be neglected. In any case, the VERITAS observations started 1.5 hours after the Whipple observations and continued with simultaneous (to Whipple) observations until the end of the night. These two instruments observed a flux enhancement by more than a factor of 5. Therefore, it is reasonable to assume that the VHE flare lasted more than 3 hours, and hence that {\it Swift}'s observations might have occurred during this bright VHE state. However, {\it Swift} did not record any substantial flux increase during MJD 54952. The flux in the energy range 0.3-2 keV is essentially compatible with the flux during previous days. In the range 2-10 keV there is a flux increase on MJD 54952, which lasted several days. However, the magnitude of the flux increase in the 2-10 keV band is about 50$\%$, which is substantially smaller than the flux increase observed by Whipple and VERITAS at the VHE domain.

{\it Fermi}-LAT operates in a survey mode, which each point of the sky is observed approximately during 30 min every 3 hours and hence sources are observed continuously on timescales down to 3 hours. However, Mrk 501 is a relatively weak source for {\it Fermi}-LAT and hence one typically needs to integrate over several days in order to have a significant detection. In these proceedings we present a light curve with the data binned in 5-day time intervals. The second-to-last time interval starts on MJD 54952 and contains the entire VHE flare, but we do not see any significant variation with respect to the previous time intervals. However, it is worth mentioning that
on MJD 54952 (day with the highest VHE flux), {\it Fermi}-LAT detects Mrk 501 in a single day with a TS larger than 25 (i.e. a significance greater than $\sim4.6\sigma$), which does not occur for the other days. The measured flux (above 300 MeV) for this day is $(3.46 \pm 2.37)~10^{-8}~~{\rm ph~ cm^{-2}~ s^{-1}} $.

\begin{figure}[!t]
  \vspace{5mm}
  \centering
  \includegraphics[width=3.in]{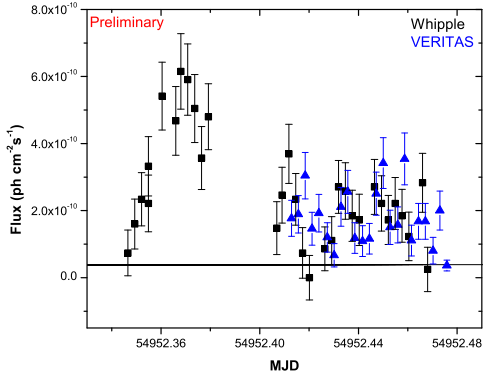}
  \caption{Whipple 10-m and VERITAS light curve (4-minute binning) for the night of the VHE flare. The solid line shows the baseline emission for the source depicted in Figure 1.} 
  \label{fig2}
 \end{figure}

Optical observations were performed as part of the Steward Observatory balzar monitoring program with the (2.3 m) Bok and the (1.54 m) Kuiper telescopes every night from MJD 54947 to MJD 54955, which includes the night of the VHE flare. Figure 3 shows the light curve, the degree of the optical linear polarization and the electric-vector position angle (EVPA) from those nights.

\begin{figure}[!t]
  \vspace{5mm}
  \centering
  \includegraphics[width=3.3in]{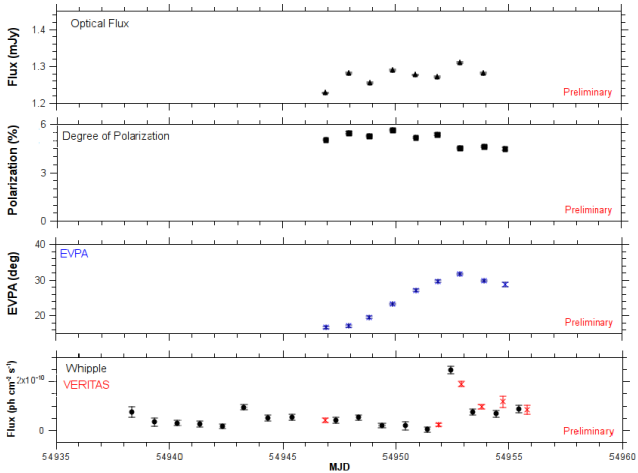}
  \caption{ Optical flux, degree of the optical linear polarization and EVPA light curves (first, second and third plots respectively) measured with the Steward Observatory, and (forth plot) the VHE light curve obtained with Whipple and VERITAS.} 
  \label{fig3a}
 \end{figure}

From MJD 54947-51, the degree of polarization was steady, and dropped a 15 $\%$ after the VHE flare. The EVPA light curve shows a continuous increase from 15 to 30 degrees in 3 days, and the rotation stops right when the large VHE flare occurs. If the two events can be physically linked, that could indicate a common origin for the optical and $\gamma$-ray emission, as already seen in\cite{lab17,lab18,lab19}  

%This behaviour is very similar to the one observed during the large outburst from PKS\,1510-089 \cite{lab17}. The correlation between the gamma-ray and optical polarized flare indicates that both emissions are produced at the same location, and have a common origin, as it occurred on several other outbursts reported recently (\cite{lab18,lab19}). As suggested in \cite{lab17}, this event could have been produced by a single blob passing through a standing shock (maybe the VLBA core) after going through the acceleration and collimation zone of the jet (where the polarization angle rotated).

\vspace{-0.5cm}
\subsection{Variability $\&$ Correlation}
\vspace{-0.2cm}
In order to quantify the flux variability in the light curves, the fractional RMS variability amplitude, $F_{var}$ \cite{lab12} was calculated as:\\
\vspace{-0.2cm}
\begin{equation}\label{ecuacion1}
\displaystyle F_{var}=\sqrt[]{\frac{S^{2}- \overline{\sigma}^{2}}{\overline{F}^{2}}}
\end{equation}

where $\overline{F}$ is the average photon flux, $S$ the standard deviation of the $N$ flux measurements and $\overline{\sigma}^{2}$ is the mean squared error.
The uncertainty of $F_{var}$ is given by:\\
\begin{equation}\label{ecuacion2}
 \displaystyle \Delta F_{var}= \sqrt[]{\left\{{\sqrt[]{\frac{1}{2N}}\cdot{}\frac{\overline{\sigma}^{2}}{\overline{F}^{2}~F_{var}}}\right\}^{2}+\left\{{\sqrt[]{\frac{\overline{\sigma}^{2}}{N}}\cdot{}\frac{1}{\overline{F}}}\right\}^{2}}
\end{equation}
% \Delta F_{var}= \frac{\overline{\sigma}^{2}}{\sqrt[]{N}~\overline{F}^{2}}\cdot{}\sqrt[]{1+\frac{1}{2~\overline{F}^{2}~F^{2}_{var}}} 

These values were calculated using a daily average for each energy band. Figure 4 shows the $F_{var}$ values obtained for all the experiments involved, except for {\it Fermi}-LAT where the result showed a negative excess variance ($\overline{\sigma}^{2} > S^{2}$), indicating low level of variability and/or slightly overestimated errors.
Essentially such a result can be interpreted as no signature for variability either because there was no variability or the instrument was not sensitive enough to detect it.

 \begin{figure}[!t]
  \vspace{5mm}
  \centering
  \includegraphics[width=2.5in]{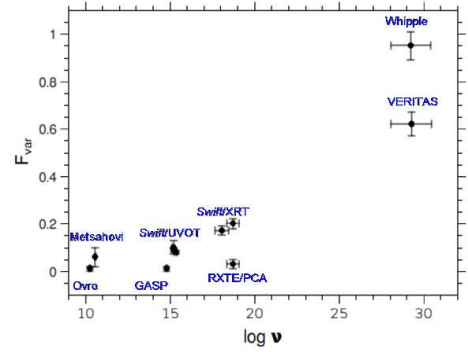}
  \caption{Fractional variability amplitude for all the instruments involved.}
  \label{fig3}
 \end{figure}

The value of $F_{var}$ is compatible with zero or very low for all the energy bands, except for the VHE range, where VERITAS observed a $F_{var}$ of 0.62$\pm$0.05 and Whipple 0.95$\pm$0.07.  This large variability in the VHE domain is clearly dominated by the large VHE flare observed on MJD 54952 and the few days after.

Historically, the X-ray and VHE $\gamma$-ray fluxes are correlated in general, yet we found that the large VHE $\gamma$-ray flare was not accompanied by a large X-ray flare. The {\it Swift}-XRT observations are not simultaneous, but are contemporaneous (7 hours late) and did not show any increase in the flux. 
The flux in the 2-10 keV band showed an increase of about 50$\%$ during the days following the VHE flare, but this flux enhancement is substantially smaller than the one seen by Whipple and VERITAS during the same days. 
At the GeV band the source remained steady with no variations during those days. To investigate the relationship between the X-ray and the VHE $\gamma$-ray, the {\it discrete correlation function} (DCF) was calculated \cite{lab15} and does not indicate a significant correlation between the two datasets.
We can conclude that this VHE flare is very probably an ``orphan'' flare (in the context of not being accompanied by an X-ray flare), like the one previously detected on 1ES\,1959+650 \cite{lab16}. Alternatively, the X-ray flare could exist, but being substantially smaller than that seen at VHE, as it occurred for PKS 2155-304 in 2006 \cite{lab20}. Further details on this subject will be given on a forthcoming publication.

\vspace{-0.5cm}
\subsection{Spectral Energy Distribution}
\vspace{-0.2cm}

The evolution of the TeV energy spectrum during the short flaring state of Mrk 501 could be important for understanding the mechanism of particle acceleration in a blazar source. The differential energy spectra of Mrk 501 (see Figure 5) with the Whipple Telescope and VERITAS were modeled for the quiescent emission and for the flare state with a simple power law in each case: \\
$ \frac{dN}{dE}= F_{\rm 0} {\times 10^{-7}}~ (E/1~\rm TeV)^{-\Gamma_{ \rm VHE}} ~~ {\rm ph~ m^{-2}~ s^{-1}~ TeV^{-1}} $
\\
where $F_{\rm 0}$ is a normalization factor and $\Gamma_{ \rm VHE}$ is the photon index.
The best-fit parameters and associated errors for the VHE data are summarized in Table \ref{table_single}.

\begin{table}[t]
\begin{center}
\scriptsize{
\begin{tabular}{c c c c c c}
\hline
& & MJD Interval &  $F_{\rm 0}$      &  $\Gamma_{ \rm VHE}$ & $\chi^{2}$/NDF \\
\hline
Whipple & very high flux  & 54952.35-54952.41 & 16.1 $\pm$ 0.4    & 2.10 $\pm$ 0.05  & 13.48/8    \\
 & high flux      & 54952.41-54955    & 5.6 $\pm$ 0.4     & 2.31 $\pm$ 0.11  & 3.10/8     \\
 & low flux       & 54936-54951       & 1.16 $\pm$ 0.09   & 2.61 $\pm$ 0.11  & 3.40/8     \\
VERITAS & high flux      & 54952-54955       & 4.17 $\pm$ 0.24   & 2.26 $\pm$ 0.06  & 6.26/5     \\
 & low flux        & 54907-55004       & 0.88 $\pm$ 0.06  & 2.48 $\pm$ 0.07  & 3.76/5     \\
\hline
\end{tabular}
}
\caption{Best-fit parameters for VHE data.}\label{table_single}
\end{center}
\end{table}

%\vspace{-0.5cm}

\begin{figure}[!t]
  \vspace{5mm}
  \centering
  \includegraphics[width=2. in]{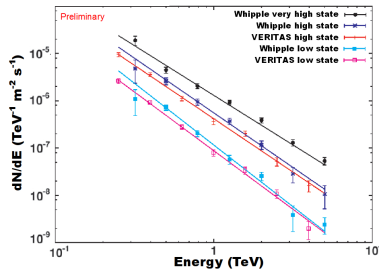}
  \caption{Time-averaged VERITAS and Whipple photon spectra of Mrk 501 for discrete flux levels (see text).}
  \label{fig5}
 \end{figure}

A marginal indication of spectral hardening with increasing flux activity is found in the TeV band. A similar trend had already been found in earlier observations in 2005 with MAGIC \cite{lab2}, when the softest photon index was 2.43$\pm$0.05 (for the low/medium state), and the hardest photon index was 2.09$\pm$0.03 (for the flare on MJD 53551). %\footnote{\footnotesize{The shape of the gamma-ray spectrum for the flare on MJD 53551 could be better fit with a log-parabola. The fit with a power-law function gave a $\chi^{2}$/NDF of 26.1/11}}

The broadband spectral results obtained for the high state (MJD 54952-55) of Mrk 501 during the three-week period (MJD 54936-56) are plotted in Figure 6. For comparison, the average spectrum for the 4.5-month multiwavelength campaign (which includes the three-week period) is also plotted \cite{lab6}. 
Figure 6 shows very clearly that the largest variation in the emitted flux occurs in the VHE domain.

  \begin{figure}[!t]
  \vspace{5mm}
  \centering
  \includegraphics[width=2.5 in]{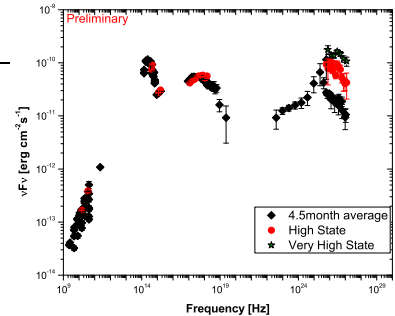}
  \caption{Spectral energy distribution of Mrk501 during the interval MJD 54952-54955 (depicted with red circles),
and the VHE spectrum measured by Whipple on MJD 54952 (depicted with green stars). For comparison purposes, the 4.5-month averaged SED from the entire campaign (extracted from \cite{lab6}) is also shown with black diamonds.}
 \label{fig6}
 \end{figure}

%Further details, SED modeling and a more extended interpretation of the results will be provided at the conference.

\vspace{-0.5cm}
\section{Acknowledgments}
\vspace{-0.3cm}
\footnotesize{
The VERITAS research is supported by grants from the US Department of Energy, the US National Science Foundation,
and the Smithsonian Institution, by NSERC in Canada, by Science Foundation Ireland, and by STFC
in the UK. We acknowledge the excellent work of the technical support staff at the FLWO and the collaborating
institutions in the construction and operation of the instrument.

The {\it Fermi}-LAT Collaboration acknowledges support from a number of agencies and institutes for both development and the operation of the LAT as well as scientific data analysis. These include NASA and DOE in the United States, CEA/Irfu and IN2P3/CNRS in France, ASI and INFN in Italy, MEXT, KEK, and JAXA in Japan, and the K.~A.~Wallenberg Foundation, the Swedish Research Council and the National Space Board in Sweden. Additional support from INAF in Italy and CNES in France for science analysis during the operations phase is also gratefully acknowledged.
}

\clearpage

\end{document}